\documentstyle[12pt,psfig,axodraw]{article} 
\newcommand{\vs}{\vspace{-0.25cm}}

\begin{document} 

\begin{center}
\large{\bf Systematic calculation of s-wave pion and kaon self-energies in
asymmetric nuclear matter}\footnote{Work supported in part by BMBF, GSI and
DFG.} 

\bigskip 

\bigskip

N. Kaiser and W. Weise\\

\bigskip

Physik Department, Technische Universit\"{a}t M\"{u}nchen,\\
    D-85747 Garching, Germany

\end{center}

\bigskip

\bigskip

\begin{abstract}
We calculate the pion self-energies in asymmetric nuclear matter in the 
two-loop approximation of chiral perturbation theory. We find three types of 
corrections beyond the well-known linear density approximation. The resulting
s-wave potential (or equivalent mass-shift) of a $\pi^-$ in the center of a 
heavy nucleus  like Pb turns out to be $U_{\pi^-}=\Delta m_{\pi^-} \simeq
14$\,MeV, about half of what is needed to form the deeply bound and narrow 
pionic atom-states recently observed at GSI. The potential for a $\pi^+$ under
the same conditions is $U_{\pi^+}=\Delta m_{\pi^+} \simeq -1$\,MeV. In the 
same way we calculate the  mass-shifts of $K^+$-mesons in symmetric nuclear
matter and neutron matter and find an increase of the $K^+$-mass by $9\%$ and
$5\%$, respectively. As a further application we give an analytical result for
the third order term in the scattering length expansion of the in-medium
self-energy of an interacting light-boson/heavy-fermion system.  
\end{abstract}


\vskip 3.5cm
The modification of pion  (or kaon) properties in dense nuclear matter is 
characterized by their density and four-momentum dependent self-energy. 
Systematic studies of  pion-nucleus scattering and pionic atom data have lead 
to a standard parametrization of the corresponding phenomenological 
pion-nucleus optical potential (for a review see \cite{batty}). On that basis 
Friedman and Soff \cite{fried} and Toki and Yamazaki \cite{toki1,toki2} 
predicted the existence of deeply  bound and surprisingly narrow pionic 
atom-states in heavy nuclei. Pionic 1s and 2p states in $^{207}$Pb and 
$^{205}$Pb have recently been observed at GSI \cite{yama,gilg} using
(d,$^3$He) transfer reactions, and there is on-going experimental activity to 
investigate more detailed aspects of such states like systematic isotope-shift 
effects. These deeply bound pionic states owe their existence to a subtle
balance between the attractive Coulomb potential and the repulsive s-wave
optical potential. The net attraction is localized at and beyond the nuclear
surface. The resulting pion wave function has little overlap with the nucleus
and this explains the observed small absorption widths, $\Gamma < 1$\,MeV. Our
aim here is to provide a reliable baseline, guided by chiral symmetry, for the
discussion of these phenomena. 

The starting point in the construction of the pion-nucleus optical potential 
is the low-density theorem \cite{ericson,dover}. It states that, to linear 
order in the proton and neutron densities, the optical potential is given by 
the respective $\pi N$-scattering threshold T-matrices  times these densities 
(see eq.(4) below). Corrections beyond the linear density approximation have
been derived by Ericson and Ericson \cite{ericson} for symmetric nuclear matter
employing analogies with optics. The correction term of \cite{ericson} can be 
interpreted as double-scattering in the nuclear medium, with the relevant 
$\pi N$-interaction parametrized through the scattering lengths.

Deeply bound pionic states are particularly sensitive to the s-wave part of
the pion-nucleus optical potential \cite{brock}. The s-wave $\pi N$-interaction
is,  on the other hand, strongly constrained by chiral low-energy theorems 
(i.e. the Tomozawa-Weinberg relation for the $\pi N$-scattering lengths). The 
tool to systematically investigate the consequences of spontaneous and explicit
chiral symmetry breaking in QCD is chiral perturbation theory. Observables are 
calculated with the help of an effective field theory formulated in terms of 
the Goldstone bosons ($\pi,K,\eta$) and the low-lying baryons. A systematic 
expansion in small external momenta and meson masses is possible. It is the 
purpose of this work to calculate the s-wave pion (and kaon) self-energies in 
asymmetric nuclear matter completely up to sixth order in the small momentum 
expansion. The results include (to that order) all corrections beyond the
linear density approximation generated by the chiral meson-baryon interactions 
together with their explicit dependence on the proton and neutron densities.

\bigskip
\bigskip

\begin{center}

\SetWidth{1.5}
  \begin{picture}(410,90)
\DashLine(5,0)(5,90){5}
\ArrowArc(25,45)(20,-180,180)
\Vertex(5,45){3}

\ArrowArc(100,45)(20,-180,180)
\DashLine(100,65)(100,90){5}
\DashLine(100,25)(100,0){5}
\Vertex(100,25){3}
\Vertex(100,65){3}

\ArrowArc(180,45)(20,-180,180)
\DashLine(180,25)(180,65){5}
\Vertex(180,25){3}
\Vertex(180,65){3}
\Vertex(160,45){3}
\DashLine(160,0)(160,90){5}

\ArrowArc(250,45)(20,-180,180)
\DashLine(250,0)(250,90){5}
\Vertex(250,25){3}
\Vertex(250,65){3}

\DashLine(310,0)(310,90){5}
\ArrowArc(330,45)(20,-90,270)
\Vertex(310,45){3}
\DashLine(310,45)(350,45){5}
\Vertex(350,45){3}

\DashLine(380,0)(380,90){5}
\ArrowArc(420,45)(20,-180,180)
\Vertex(380,45){3}
\DashLine(380,45)(420,65){5}
\DashLine(380,45)(420,25){5}
\Vertex(420,25){3}
\Vertex(420,65){3}

\end{picture}
\end{center}

\medskip
\noindent
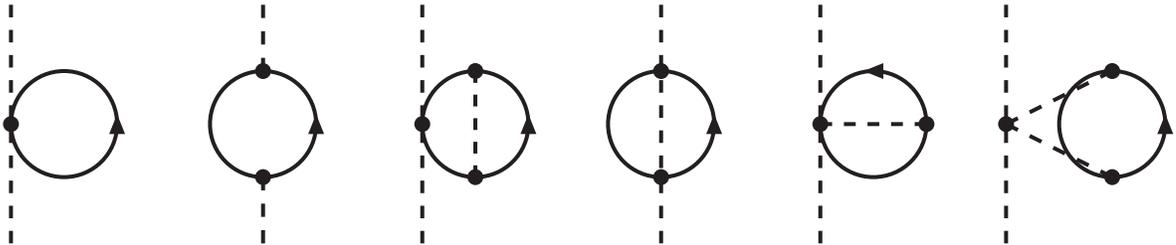
{\it Fig.1: Some in-medium self-energy diagrams up to two loops. Solid and 
dashed lines represent nucleons and pions, respectively. The combinatoric 
factor of the last diagram is 1/2.}
\vskip 0.7cm

From the start we will restrict the calculation of the in-medium self-energies 
to on-shell pions (and kaons) with zero three-momentum, i.e. $q^\mu=(m_{\pi,K},
\vec 0\,)$. The only new ingredient to perform calculations at finite baryon
density (instead  of scattering processes in the vacuum) is the
particle-hole or in-medium nucleon propagator. It expresses the fact
that the ground-state has changed from an empty vacuum to a filled Fermi-sea of
nucleons. In the heavy baryon limit the in-medium nucleon propagator reads
\begin{equation} {i \over p_0+i\epsilon}-2\pi \delta(p_0) \bigg[{1+\tau_3\over
2}\, \theta(k_p-|\vec p\,|)+{1-\tau_3\over 2}\,\theta(k_n-|\vec p\,|)\bigg] \,,
\end{equation} 
with the energy $p_0$ counted modulo the large nucleon mass $M$ and $k_{p,n}$
are the Fermi-momenta of the protons and neutrons, related to their densities 
$\rho_{p,n}=k_{p,n}^3/3\pi^2$ in the usual way. With the help of the projection
operators $(1\pm \tau_3)/2$ one can easily treat the case of different proton
and neutron densities (i.e. isospin-asymmetric nuclear matter). Note that the 
in-medium  propagator eq.(1) splits additively into the vacuum propagator and a
medium-insertion. This allows to organize the diagrammatic calculation 
according to the number of medium insertions. Diagrams with no medium insertion
just renormalize the meson mass to its (measured) vacuum value. Diagrams with
one medium insertion (such as the first one in fig.\,1) produce exactly one
factor of density $\rho_p$ or $\rho_n$ and thus contribute to the linear
density approximation. Corrections beyond the linear density approximation 
arise only from diagrams with (at least) two medium insertions (see fig.\,1).
Their evaluation leads to an integral over two Fermi-spheres of 
different radii which can be reduced to a simple one-parameter integral by the
following formula, 
\begin{equation} \int {d^3p_1 d^3p_2 \over (2\pi)^6}\, G(|\vec p_1-\vec p_2|)\,
\theta(k_1-|\vec p_1|)\,\theta(k_2-|\vec p_2|) = {(k_1+k_2)^6 \over 192
\pi^4 } \int_0^1 dx\, x \chi(x) \, G(x(k_1+k_2)) \,, \end{equation}
with the weight-function 
\begin{equation} \chi(x) = 2x(1-r)^3 \, \theta(r-x)+(1-x)^2(x^2+2x-3r^2) \, 
\theta(x-r) \,, \qquad r= {|k_1-k_2|\over k_1+k_2} \,. \end{equation} 
Furthermore, working at $\vec q=\vec 0$ simplifies the calculation considerably
since all two-loop diagrams in which the in- or outgoing pion couples directly
to the nucleon vanish identically ($\vec \sigma \cdot \vec q=0$). The third
diagram in fig.\,1, with three nucleon propagators, requires some special 
attention. Here, one encounters the square of the in-medium propagator eq.(1)
whose medium-insertion turns out to be proportional to the derivative of a 
delta-function $\delta'(p_0)$. This together with the fact that the internal 
pion-propagator is an even function of $p_0$, implies that there is no 
correction beyond the linear density approximation from this particular 
diagram.

We are now in the position to present the complete two-loop result for the pion
self-energies in asymmetric nuclear matter.  In the case of a negatively 
charged $\pi^-$ it reads: 
\begin{equation} \Pi^-(k_p,k_n) =  {k_n^3\over 3\pi^2}\Big[T^-_{\pi N}
-T^+_{\pi N} \Big] - {k_p^3 \over 3\pi^2}\Big[T^-_{\pi N}+T^+_{\pi N} \Big] 
+\Pi^-_{rel}(k_p,k_n) +\Pi^-_{cor}(k_p,k_n)\,. \end{equation}
The first two terms represent the linear density approximation with
$T^{\mp}_{\pi N}=4\pi (1+m_\pi/M) a^\mp$, the isospin-odd and isospin-even 
$\pi N$-scattering threshold T-matrices calculated to 1-loop order in chiral
perturbation theory \cite{pin1,pin2}, 
\begin{equation} T^-_{\pi N} ={m_\pi \over 2f_\pi^2} + {g_A^2 m_\pi^3 \over 8 
M^2 f_\pi^2}+ {m_\pi^3 \over (4\pi f_\pi^2)^2 } \Big[ 1-2\ln{m_\pi \over
\lambda} +b(\lambda)\Big] +{\cal O}(m_\pi^5) \,, \end{equation}
\begin{equation} T^+_{\pi N} 
= {m_\pi^2 \over f_\pi^2}\Big[2C-{g_A^2 \over 4M} \Big] + {3g_A^2 
m_\pi^3 \over 64\pi f_\pi^4}+ {\cal O}(m_\pi^4) \,. \end{equation}
Here, $f_\pi=92.4\,$ MeV is the weak pion decay constant and $g_A=1.27$ the
nucleon axial-vector coupling constant. The combinations of low-energy 
constants $C=c_2+c_3-2c_1$ and $b(\lambda)$ have been discussed in 
detail in refs.\cite{pin1,pin2} and the vanishing of the $m_\pi^4$-term for 
$T^-_{\pi N}$ has been proven in ref.\cite{pin2}. Experimental values for 
$T^\mp_{\pi N}$ have been extracted from the shift and width of the 1s level 
in pionic hydrogen by the PSI group \cite{psi}, with the final result  
\begin{equation} T^-_{\pi N} = (1.847\pm 0.086) \,{\rm fm}\,, \qquad
T^+_{\pi N} = (-0.045\mp 0.088) \,{\rm fm}\,. \end{equation}
It is important to note that the pion-loop correction closes the gap between 
the Tomozawa-Weinberg prediction $m_\pi/2f_\pi^2 = 1.61$\,fm and the empirical
value of $T_{\pi N}^-$. In the case of $T^+_{\pi N}$ no accurate chiral
prediction is possible since the next-to-leading order pion loop term $3g_A^2
m_\pi^3/64\pi  f_\pi^4=0.18$\,fm is still large in comparison to the
(anomalously) small empirical value of $T^+_{\pi N}$ (which is actually
compatible with zero).    

The last two terms in eq.(4) comprise the corrections beyond the linear density
approximation. The first term
\begin{equation} \Pi^-_{rel}(k_p,k_n)={g_A^2 m_\pi \over 10(M\pi
f_\pi)^2} (k_p^5-k_n^5) \,, \end{equation}
is a small relativistic correction from the second diagram in fig.\,1. It
originates from the energy-dependent pseudovector $\pi N$ Born amplitude in
forward direction, integrated over the Fermi-sphere. More 
interesting are the contributions from the last three diagrams in fig.\,1, with
two medium insertions. They describe scattering processes in which two 
nucleons in the Fermi-sea participate, with the result 
\begin{eqnarray} \Pi^-_{cor}(k_p,k_n) &=& {4m_\pi^2 \over (4\pi f_\pi)^4}
\bigg\{ 2(k_p^2+k_n^2)^2 +2 k_p k_n(k_p-k_n)^2 +(k_p^2-k_n^2)^2
\ln{|k_p-k_n|\over k_p+k_n} \bigg\} \nonumber \\ && +{g_A^2 m_\pi^2 \over (4\pi
f_\pi)^4 } \bigg\{ 2k_p^2(m_\pi^2-2k_p^2)+ 2k_n^2(m_\pi^2-2k_n^2) \nonumber \\
& & +8k_p^3 m_\pi\arctan{2k_p \over m_\pi} +8k_n^3 m_\pi\arctan{2k_n \over 
m_\pi} \nonumber \\ && -m_\pi^2 \Big( {m_\pi^2 \over 2} +4k_p^2 \Big)\,\ln\Big(
1+{4k_p^2 \over m_\pi^2} \Big) -m_\pi^2 \Big( {m_\pi^2 \over 2} +4k_n^2 \Big)\,
\ln\Big( 1+ {4k_n^2 \over m_\pi^2} \Big) \bigg\} \,. \end{eqnarray}
The first line in eq.(9) is the Ericson-Ericson double scattering term 
\cite{ericson} generalized to asymmetric nuclear matter ($k_p\ne k_n$). The
term proportional to $g_A^2$ is new. It represents the effect induced by
$\pi\pi$-interaction with two virtual pions being absorbed on the nucleons
in the Fermi-sea. Note that the off-shell $4\pi$-vertex is not uniquely
defined by the chiral Lagrangian and therefore one should view the last two 
diagrams in fig.\,1 always as one entity. Such ambiguities hint at 
the fact that an off-shell in-medium self-energy itself is not a unique 
concept in field  theory. Interestingly, the two-nucleon correlation term
$\Pi^-_{cor}(k_p,k_n)$ is an even function of the proton-neutron asymmetry.   

Next, we turn to the $\pi^+$ self-energy in asymmetric nuclear matter. It can
be easily obtained from the $\pi^-$ one by an isospin transformation
which exchanges the role of protons and neutrons, 
\begin{equation} \Pi^+(k_p,k_n) =\Pi^-(k_n,k_p) \,. \end{equation}
For the sake of completeness  we also give the result for the (neutral) 
$\pi^0$ in-medium self-energy,
\begin{equation} \Pi^0(k_p,k_n) = -T^+_{\pi N} {k_p^3+k_n^3 \over 
3\pi^2}+ \Pi^0_{cor}(k_p,k_n)\,, \end{equation}
with the corresponding two-nucleon correlation term,
\begin{eqnarray} \Pi^0_{cor}(k_p,k_n)&=& {m_\pi^2 \over(2\pi f_\pi)^4}
\bigg\{ k_pk_n(k_p^2+k_n^2) +{1\over 2}(k_p^2-k_n^2)^2 \ln{|k_p-k_n|\over 
k_p+k_n} \bigg\} \nonumber \\ &&+ {g_A^2 m_\pi^2 \over (4\pi f_\pi)^4 } 
\bigg\{ 2k_p^2 (2k_p^2-m_\pi^2)+ 2k_n^2(2k_n^2-m_\pi^2)-8k_p^3 m_\pi\arctan
{2k_p \over m_\pi} \nonumber \\ && -8k_n^3 m_\pi\arctan{2k_n \over m_\pi} 
+m_\pi^2 \Big({m_\pi^2 \over 2} +4k_p^2 \Big)\,\ln\Big( 1+{4k_p^2\over m_\pi^2}
\Big)  \nonumber \\ && + m_\pi^2 \Big( {m_\pi^2 \over 2} +4k_n^2 \Big)
\ln\Big( 1+ {4k_n^2 \over m_\pi^2} \Big) +8k_pk_n(m_\pi^2-k_p^2-k_n^2) 
\nonumber \\ && +16(k_p^3 +k_n^3) m_\pi \arctan{k_p+k_n\over m_\pi}  
-16|k_p^3-k_n^3| m_\pi  \arctan{|k_p-k_n|\over m_\pi} \nonumber \\ &&
+2 \Big[(k_p^2-k_n^2)^2 -4m_\pi^2 (k_p^2+k_n^2)-m_\pi^4 \Big]
\ln{m_\pi^2+(k_p+k_n)^2 \over m_\pi^2+ (k_p-k_n)^2} \bigg\} \,. \end{eqnarray} 
The differences in comparison with the charged pion case can be traced back to
certain different isospin factors. 

The self-energies are related to the threshold s-wave 
optical potentials $U^{\mp0}(k_p,k_n)$ by  $\Pi^{\mp0}(k_p,k_n) = 2m_\pi 
U^{\mp0}(k_p,k_n)$, and we can identify in-medium pion mass-shifts, 
$\Delta m_{\pi^{\mp0}}= U^{\mp0}(k_p,k_n)$ in order to give a physical
interpretation to these potentials. For a numerical example we consider 
asymmetric nuclear matter of density $(k_p^3+k_n^3)/3\pi^2 = 0.165 
$\,fm$^{-3}$ with a neutron-to-proton ratio  $N/Z=(k_n/k_p)^3=1.5$ as it is 
typical for a heavy nucleus like Pb. This implies proton and neutron
Fermi-momenta of $k_p=246.7$\,MeV and $k_n=282.4$\,MeV.  The in-medium pion
mass-shifts $\Delta m_{\pi^{\mp0}}$ are found in such a dense environment as
 
\begin{equation} \Delta m_{\pi^-}=(9.5-1.0+6.0-0.7)~{\rm MeV}= 13.8~{\rm MeV}
\,,\end{equation}
\begin{equation} \Delta m_{\pi^+}=(-7.5+1.0+6.0-0.7)~{\rm MeV} = -1.2~{\rm MeV}
\,,\end{equation}
\begin{equation} \Delta m_{\pi^0}=(1.0+0.0+5.7-0.6)~{\rm MeV}=6.1~{\rm MeV}\,.
\end{equation}
Here, the first number in brackets gives the contribution from the linear 
density  approximation using the central PSI-values for $T^\mp_{\pi N}$ (see
eq.(7)). The second number corresponds to the small relativistic correction
eq.(8) (zero for $\pi^0$), the third number gives the double scattering 
contribution and the fourth number corresponds to the new $g_A^2$-terms induced
by $\pi\pi$-interaction. Note that the splitting of $\pi^-$ and $\pi^+$ in
asymmetric nuclear matter comes almost entirely from the leading order
$T^-_{\pi N}$-term driven by the Tomozawa-Weinberg low-energy theorem. This
splitting is absent in symmetric nuclear matter. Clearly the Ericson-Ericson 
double scattering term \cite{ericson} is the dominant correction beyond the 
low density approximation and it is almost equal for charged and neutral pions
even in asymmetric nuclear matter. The effect induced by $\pi\pi$-scattering 
reduced this by about $10\%$. Both these two-nucleon correlation effects show 
up in a calculation of the pion-deuteron scattering length \cite{beane}. There,
the $g_A^2$-term contributes only at the $3-4\%$ level due to the much smaller
densities relevant in the deuteron. 
    
A repulsive mass-shift $\Delta m_{\pi^-}\simeq 14$\,MeV is still too small (by
about a factor of two)  to form the deeply bound pionic atom-states in Pb as
narrow as they are observed. We have therefore calculated some contributions of
seventh order in the small momenta $m_\pi,k_p,k_n$. For example,
continuing the $1/M$-expansion of the second diagram in fig.\,1 one obtains    
\begin{equation}  \Pi^-_{rel}(k_p,k_n)= -{3 g_A^2 m_\pi^2 (k_p^5+k_n^5) 
\over 40 \pi^2 M^3f_\pi^2 } \,, \end{equation}
which results in an additional small downward mass-shift of $\Delta m_{\pi^-}
=-0.3$~MeV. In the dominant double scattering diagram (fourth graph in fig.\,1)
one can either correct for the nucleon kinetic energies in the pion-propagator
or replace the upper or lower leading order vertex by a second order vertex
proportional to $C-g_A^2/8M$. Both types of corrections give together,  
\begin{equation} \Pi^-_{cor}(k_p,k_n)= {m_\pi^3 (k_n^2-k_p^2) \over M 
(2\pi f_\pi )^4 } \bigg\{ k_p k_n +{1\over 2}(k_p^2+k_n^2) \Big[\ln{|k_p-k_n|
\over k_p+k_n} +g_A^2 -8MC\Big]\bigg\} \,,
\end{equation} 
and numerically they cancel each other, $\Delta m_{\pi^-} = (-0.2+0.2)$~MeV.

In phenomenological calculations \cite{gal} it has been common practice to
introduce a term, $-16\pi B_0\,\rho_p\rho_n$, in the $\pi^-$ self-energy and to
choose Re\,$B_0$ such that the missing repulsion is accounted for. A term of
this form would show up at three-loop order in connection with pion-absorption,
$\pi NN \to NN$. This process involves large momenta of order $\sqrt{Mm_\pi}$ 
which are difficult to cope with in the counting scheme of chiral perturbation 
theory (see also the discussion of the inverse process $NN\to NN\pi$ at
threshold in ref.\cite{nnx}). A large negative value of Re\,$B_0$ \cite{batty}
would be at odds with theoretical many-body calculations \cite{riska}, though
within considerable uncertainties \cite{holinde}. An alternative way to
generate the missing s-wave repulsion could be the in-medium renormalization of
the pion decay  constant $f_\pi$ \cite{weise}.

The two-loop ChPT calculation of the pion self-energy in asymmetric nuclear
matter can be easily taken over to the case of kaons ($K^+,K^0$). The 
mass-shifts of anti-kaons ($K^-,\bar K^0)$, on the other hand, are strongly 
affected by the (non-perturbative) in-medium dynamics of the subthreshold 
$\Lambda(1405)$-resonance \cite{waas}. At two-loop order  the $K^+$ 
self-energy in asymmetric nuclear matter reads
\begin{eqnarray} \Pi_{K^+}(k_p,k_n) &=& -{k_n^3\over 6\pi^2}\,T_{KN}^{(0)}
-{2k_p^3+k_n^3 \over 6\pi^2}\,T_{KN}^{(1)}\nonumber \\ &&+ {2m_K^2
\over (4\pi f_K)^4} \bigg\{ 16k_p^4+4k_n^4+2k_p k_n(k_p^2+k_n^2)
+(k_p^2-k_n^2)^2 \ln {|k_p-k_n|\over k_p+k_n} \bigg\} \,,\end{eqnarray}
with $T^{(0,1)}_{KN}=4\pi(1+m_K/M)a^{(0,1)}_{KN}$ the isospin 0 and isospin 1 
$KN$-scattering threshold T-matrices. The empirical values $T_{KN}^{(0)}=
0.4$~fm and $T_{KN}^{(1)}= -6.3$~fm stem from a dispersion relation analysis 
of $KN$ and $\bar KN$-scattering by Martin \cite{martin}. The expression in 
the second line of eq.(19) derives from the double scattering graph (fourth 
diagram in  fig.\,1) using chiral $KN$-interaction vertices. Here we have 
introduced the weak kaon decay constant $f_K=113$\,MeV in order to have a
more realistic leading order value for $T^{(1)}_{KN}=-m_K/f_K^2=-7.6\,$fm. Of
course, we have also evaluated the corrections beyond the linear density 
approximation which come from the $1/M$-expansion of the pseudovector $KN$
Born-amplitude and from the diagrams with virtual $(\pi\pi\,, \pi^0\eta\,,
\eta\eta)$-exchange between the kaon and two nucleons in the Fermi-sea. We
found that these lead to negligibly small kaon mass-shift of $0.6$\,MeV or
less. Therefore we have omitted the corresponding expressions in eq.(19).

The $K^+$ mass-shift in symmetric nuclear matter of density $0.165$~fm$^{-3}$ 
is  $\Delta m_K= (30+15)$~MeV = $45$~MeV according to eq.(19), with the
first and second number coming from the linear density approximation and the 
double scattering term, respectively. In pure neutron matter at the same 
density the $K^+$ mass-shift is substantially smaller,  $\Delta m_K=
(19+6)$~MeV = $25$~MeV.  The coupled channel calculation of \cite{waas} has
found very similar results. Again, the $K^0$ self-energy can be immediately
obtained by an isospin transformation as $\Pi_{K^0}(k_p,k_n)=\Pi_{K^+}(k_n,k_p)
$. In this context, we have to comment on the "model independent" result for
the $K^+$ mass-shift of Lutz \cite{lutz}. For symmetric matter ($k_p=k_n)$ (and
to the order in $m_K/M$ considered here) we agree with the real part of his
double scattering term. The claimed imaginary part (implying a $K^+$ decay
width of $\Gamma_{K^+}=4\,$MeV) is  however wrong. For a $K^+$ at rest all 
on-shell scattering processes to change any of its quantum numbers are
Pauli-blocked.  

\bigskip
\bigskip

\begin{center}

\SetWidth{1.5}
  \begin{picture}(410,90)
\DashLine(5,0)(5,90){5}
\ArrowArc(25,45)(20,-180,180)
\Vertex(5,45){3}

\ArrowArc(100,45)(20,-180,180)
\DashLine(100,0)(100,90){5}
\Vertex(100,25){3}
\Vertex(100,65){3}

\ArrowArc(170,45)(20,-180,180)
\DashLine(180,0)(180,28){5}
\DashLine(180,62)(180,90){5}
\Vertex(180,28){3}
\Vertex(180,62){3}
\DashLine(180,28)(150,45){5}
\DashLine(180,62)(150,45){5}
\Vertex(150,45){3}

\DashLine(230,0)(230,28){5}
\ArrowArc(240,45)(20,0,360)
\Vertex(230,28){3}
\DashLine(230,62)(230,90){5}
\DashLine(230,62)(260,45){5}
\DashLine(260,45)(230,28){5}
\Vertex(230,62){3}
\Vertex(260,45){3}

\end{picture}
\end{center}

\medskip
\noindent
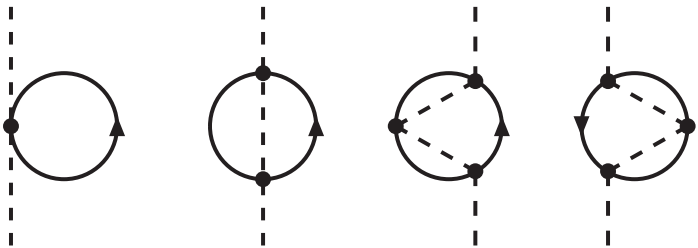
{\it Fig.2: In-medium self-energy diagrams up to third order in the scattering
length $a$.} 

\bigskip
       
As a side remark and a further application of effective field theory in the
same context, we consider the self-energy a light scalar boson moving in a 
medium consisting of heavy spin-1/2 fermions. The boson-fermion interaction is 
assumed to be completely described by the scattering length $a$, i.e. a local 
contact interaction of the form ${\cal L}_{int} = 2\pi \, a\, \bar HH \phi^2$. 
From the diagrams in fig.\,2 one can compute the in-medium self-energy 
$\Pi(a,k_f)$ up to third order in the scattering length $a$, with the 
analytical result  
\begin{equation} \Pi(a,k_f)= -{4\over 3\pi}\, a\,k_f^3+{2\over \pi^2}\, 
a^2\,k_f^4-{8(15-\pi^2)\over 45\pi^3}\, a^3\,k_f^5\,, \qquad \quad 
\rho ={k_f^3 \over 3\pi^2} \,. \end{equation}
The new triple scattering term (which goes beyond the work of 
ref.\cite{ericson}) originates from the two topologically distinct three-loop 
diagrams in fig.\,2, either with three medium insertions or with two medium 
insertions at the lower and upper fermion-propagator. The latter combination 
is not a renormalization of the $a^2k_f^4$ double scattering term. Eq.(20) 
cannot be directly applied to the $\pi N$ and $KN$ systems due to the 
additional isospin quantum number and the host of momentum dependent chiral
meson-baryon vertices. Nevertheless, the result eq.(20) may be useful for
certain condensed matter systems.

In summary, we have used chiral perturbation theory to calculate the complete 
pion and kaon self-energies in asymmetric nuclear matter up to two-loop order. 
We have found new corrections beyond the linear density approximation together 
with their explicit dependence on the proton and neutron Fermi-momenta. To that
order the  $\pi^-$ mass-shift in Pb is about half of what is needed to form the
observed deeply bound and narrow pionic states. The present result can also be
of some guidance for the analysis of $\pi^+$ electroproduction off $^3$He
\cite{richter}.  For the $K^+$ mass-shift we confirm earlier results of model 
calculations.

\bigskip 
\subsection*{Acknowledgement}
We are grateful to P. Kienle, A. Richter and J. Wambach for stimulating
discussions.

\end{document}